\documentclass{article}
\usepackage{graphicx}
\usepackage{enumerate}
\usepackage[square]{natbib}
\usepackage{authblk} 
\usepackage{amsthm} 
\usepackage{amssymb} 
\usepackage{caption}

\newtheorem{mydef}{Definition} 
\newtheorem{theorem}{Theorem} 

\date{\today}

\begin{document}

\title{A classification scheme for interpretations of Quantum Mechanics}
\author[1]{Gerd Ch.Krizek}
\affil{University of Vienna - Quantum particle group}
\affil{UAS Technikum Wien - Department of Applied Mathematics and Sciences}


\maketitle

%

\begin{abstract}
\noindent 
A physical theory consists of the mathematical formalism and an interpretation, which contains the definition of symbols, measurement assignments, concepts and principles, and an ontology. We present a scheme to classify these different levels of a physical theory and apply it to Newtonian Mechanics and the interpretations of Quantum Mechanics. We show that this classification scheme is embedded in the methodology of philosophy of science. With this scheme, different interpretations of Quantum Mechanics can be compared concerning the formalism and the used conceptions, and it serves as a guidance to identify ontological entities and reality conceptions in the different interpretations. Inspired by the commitments on the ontological level, we propose two heuristics concerning ontological statements. 

\end{abstract}

\newpage

\section{Introduction}
\parindent 0cm       

Quantum Mechanics is undoubtedly one of the most successful theories in physics, but its foundations and interpretations are still a heavily disputed topic. Quantum Mechanics allows for various interpretations, but it is surprising how strong the differences can be concerning the ontological commitments of these interpretations. \newline

The scope of this work is to propose a classification scheme that will provide a methodology to analyse the differences of several interpretational approaches in the mathematical model, the interpretation of the respective mathematical model, the used concepts, and the ontology. It provides a number of levels to distinguish between these aspects. \newline

The scheme itself claims not to reflect any ontological aspect on the nature of theories; the levels could be defined differently and the boundaries between the levels could be chosen differently, but the proposed scheme is valuable to contrast the content of different approaches in the interpretations of Quantum Mechanics. \newline

We present the relation between the classification scheme and approaches to the structure of scientific theories in philosophy of science. A physical theory rests also on preconceptions and premises. To illustrate this, a brief overview of philosophical stances, in context with the conception and structure of physical theories, is presented. \newline

Connected to the ontological level of the classification scheme, two arguments are proposed that refer to the ontologies of theories and their interrelation. \newline

By means of an application on classical particle mechanics and on different interpretations of Quantum Mechanics, we demonstrate the merit of the scheme. 

\newpage

\section{A classification scheme}

Interpretational debates as part of the scientific method are as old as science itself. But no interpretational debate in physics attracted more attention than the debate in the interpretations of Quantum Mechanics \cite[]{landsman2006champions}. The reason is certainly that it is more than just a debate on different models about a physical situation. It is a debate with pervasive philosophical implications. \newline

The field of interpretations of Quantum Mechanics seems to be growing continuously, without reconsideration. A quote by David Mermin reflects this situation \cite[]{mermin2012commentary}: 

\begin{quote}
\textit{``Quantum mechanics is the most useful and powerful theory physicists have ever devised. Yet today, nearly 90 years after its formulation, disagreement about the meaning of the theory is stronger than ever. New interpretations appear every year. None ever disappear.''}
\end{quote}

To provide a better overview on the interpretations of Quantum Mechanics, we propose a classification scheme that categorizes the structure of a physical theory, particular an interpretation of Quantum Mechanics.

A physical theory consists roughly of the mathematical formalism and an interpretation, which contains the definition of symbols, theoretical terms, measurement assignments, concepts and principles, and an ontology. We refine this coarse breakdown into a four-level classification.

\begin{mydef}
Level 1 of a physical theory is described by a set of mathematical quantities. These mathematical quantities are in relation to each other by mathematical laws. 
\end{mydef}

\begin{mydef}
Level 2 of a physical theory is described by a set of measurement assignments to the mathematical quantities of Level 1. It describes how the defined quantities in the mathematical description of the theory are related to sensations and experience. It relates mathematical objects to theoretical terms.  
\end{mydef}


\begin{mydef}
Level 3 of a physical theory is described by a number of concepts and principles that are related to the mathematical description of the theory. These principles can be formulated as statements or logical expressions.
\end{mydef}

\begin{mydef}
Level 4 of a physical theory contains the ontology of the theory and the metaphysical entities or conceptions that are introduced by the theory. 
\end{mydef}

\newpage

We make use of the following figurative elements to visualize the classification scheme: 

\begin{table}[h]
\centering

\label{SchemeTheory}
\begin{tabular}{cc}

\begin{tabular}[c]{@{}c@{}}\\ \includegraphics[width=0.025\textwidth, height=8px]{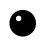} \\ \\ \end{tabular}  & \begin{tabular}[c]{@{}c@{}}  \\ Physical quantity \\ \end{tabular} \\ 
\begin{tabular}[c]{@{}c@{}}\\ \includegraphics[width=0.1\textwidth, height=12px]{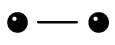} \\ \\ \end{tabular} & \begin{tabular}[c]{@{}c@{}}  \\ Law between physical quantities \\ \end{tabular} \\ 
\begin{tabular}[c]{@{}c@{}}\\\includegraphics[width=0.1\textwidth, height=40px]{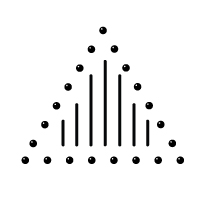} \\ \\ \end{tabular} & \begin{tabular}[c]{@{}c@{}} \\ Concept connecting laws and physical quantities \\ \end{tabular}                                                                                 \\ 
\begin{tabular}[c]{@{}c@{}} \includegraphics[width=0.15\textwidth, height=50px]{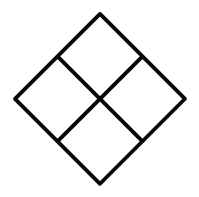} \\ \\ \end{tabular} & \begin{tabular}[c]{@{}c@{}} Elements of objective reality \\ \end{tabular}                      \\ 
\end{tabular}
\caption{Figurative elements used in the classification scheme}
\end{table}

By dots, we depict a physical quantity, which can be the property of a physical object or any other physical measure that is introduced by the theory. On Level 1 of the classification scheme, a physical quantity is basically a mathematical symbol. These dots are in relation to each other by mathematical relations, the laws between the physical quantities. On the first level, these are the mathematical laws that are fulfilled by the mathematical symbols representing the physical quantities. \newline

On the second leyel, the physical quantities receive names and relation to empirical results, the measurement assignments. By that, the pure mathematical formalism becomes a physical law. The physical quantities and the laws involve conceptions that are set up by theoretical terms and governed by principles. These are depicted by the areas spanned by some physical quantities and laws, and belong to the Level 3 of the classification scheme. \newline

On the ontological level, Level 4, there are entities that are a priori unknown. A physical theory usually provides a statement on the ontology, statements that give an insight on the metaphysical reality behind the empirical content. These statements involve usually entities that exist independently of our perception; we denote them as elements of objective reality and depict them by a diamond. 

\newpage

Applying these figurative elements to the classification scheme, we can give an overview on the scheme: \newline

\begin{table}[h]
\centering

\label{Scheme}
\begin{tabular}{|l|c|c|}
\hline
1 & \begin{tabular}[c]{@{}c@{}}\\ \includegraphics[width=0.15\textwidth, height=50px]{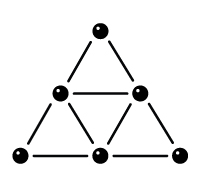} \\ \\ \end{tabular}  & \begin{tabular}[c]{@{}c@{}}\\ Mathematical laws \\ between mathematical symbols \\ \\ \end{tabular}                 \\ \hline
2 & \begin{tabular}[c]{@{}c@{}}\\ \includegraphics[width=0.15\textwidth, height=50px]{1} \\ \\ \end{tabular} & \begin{tabular}[c]{@{}c@{}}  \\ Interpretation physical quantities \\ Relation physical quantities - Measurements \\ Measurement laws \\ \\ \end{tabular} \\ \hline
3 & \begin{tabular}[c]{@{}c@{}}\\ \includegraphics[width=0.15\textwidth, height=50px]{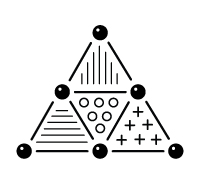} \\ \\ \end{tabular} & \begin{tabular}[c]{@{}c@{}} \\ Concepts and principles  \\ \\ \end{tabular}                                                                                 \\ \hline
4 & \begin{tabular}[c]{@{}c@{}}\\ \includegraphics[width=0.10\textwidth, height=40px]{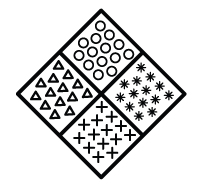} \\ \\ \end{tabular} & \begin{tabular}[c]{@{}c@{}}\\ Ontology - beables - \\ elements of reality \\ \\ \end{tabular}                      \\ \hline
\end{tabular}
\caption{Classification scheme}
\end{table}

\newpage

\section{The structure of scientific theories in the framework of philosophy of science}


Philosophy of science forms a framework for the scientific method, which evolved out of early philosophy. Philosophy was and is pervaded by dualities. One of those foundational dualities is the contrast between Experience and Reason, or as \cite[]{einstein1934method} puts it into words: 

\begin{quote}
\textit{``I want now to glance for a moment at the development of the theoretical method, and while doing so especially to observe the relation of pure theory to the totality of the data of experience. Here is the eternal antithesis of the two inseparable constituents of human knowledge, Experience and Reason, within the sphere
of physics.''}
\end{quote}

Another foundational duality is set up by the realist and positivist positions. In the framework of these dualities, the views of scientists on the construction of scientific theories evolved in the last century, in a complex way, to a diversity of positions. We present an overview on those positions that are relevant to the presented classification scheme, because these positions provide a way to give structure to scientific theories. We present the positions and their relation to the classification scheme.

\subsection{Einstein's model of layers}

Einstein presented his view on the structure of scientific theories in his work \cite[]{einstein1936physik}: 

\begin{quote}
\textit{``Ziel der Wissenschaft ist erstens die m\"oglichst vollst\"andige begriffliche Erfassung und Verkn\"upfung der Sinneserlebnisse in ihrer ganzen Mannigfaltigkeit, zweitens aber die Erreichung dieses Zieles unter Verwendung eines Minimums yon prim\"aren Begriffen und Relationen (Streben nach m\"oglichster logischer
Einheitlichkeit des Weltbildes bezw. logischer Einfachheit seiner Grundlagen).
Die Wissenschaft braucht die ganze Mannigfaltigkeit der prim\"aren, d.h. unmittelbar mit Sinneserlebnissen verkn\"upften Begriffe sowie der sie verkn\"upfenden S\"atze. In ihrem ersten Entwicklungsstadium enth\"alt sie nichts weiter. 
Auch das Denken des Alltags begn\"ugt sich im grossen Ganzen mit dieser Stufe. Diese kann aber einen wirklich wissenschaftlich eingestellten
Geist nicht befriedigen, da die so gewinnbare Gesamtheit von Begriffen und Relationen der logischen Einheitlichkeit v\"o11ig entbehrt. Um diesem Mangel abzuhelfen, erfindet man ein begriffs- und relations\"armeres System, welches
die prim\"aren Begriffe und Relationen der "ersten Schicht" als
logisch abgeleitete Begriffe und Relationen enth\"alt. Dieses neue "sekund\"are System" erkauft die gewonnene h\"ohere logische Einheitlichkeit mit dem Umst\"ande, dass seine an den Anfang gestellten Begriffe (Begriffe der zweiten Schicht) nicht
mehr unmittelbar mit Komplexen von Sinneserlebnissen verbunden sind.
... So geht es fort, bis wir zu einem System von denkbar gr\"osster Einheitlichkeit und Begriffsarmut der logischen Grundlagen gelangt sind, das mit der Beschaffenheit des sinnlich Gegebenen vereinbar ist.''}
\end{quote}

\begin{quote}
\textit{``The goal of science is in the first place the complete conceptual acquisition and connection of the sensations in all its manifold aspects, secondly the attainment of this goal with a minimum of primary terms and relations. (ambition for a optimized logical unity of the world view, respectively logical simplicity of its foundations).
Science needs the whole manifold of primary terms, i.e. which are directly connected to sensations, and the statements connecting them. In its first development stage it contains nothing more. 
Common sense thinking is satisfied with this stage. Though this stage cannot satisfy a truly scientifically oriented mind, since the resulted wholeness of conceptions and relations totally lacks in logical unity. To resolve this insufficiency one invents a system more economical in its use of conception and relations, which integrates the conceptions and relations of the "first layer" as logically derivable. This new "secondary system" gains the higher logical unity by the circumstance that its initial conceptions (conceptions of the second layer) are not directly related to the sensations any more.
...
And so it goes forth, until we reached at a system of at most thinkable unity and parsimony of logical foundations, which is in agreement with the sensations.\footnote{Translation by the author}''}
\end{quote}

The difference to the classification scheme presented before is that Einstein's model of layers refers only to Level 1 and 2 of our classification scheme  presented initially. The development of higher layers by unification and definition of new theoretical terms and concepts evolves the layer concept to the Level 3 of the classification scheme, to concepts and principles. Einstein's layer model pictures the evolution of a physical theory from its first conceptions out of common sense, pure sensations and experience, to a physical theory with a maximum of parsimony in its used elements. \newline

\begin{figure}[h]
\centering

\includegraphics[width=0.7\textwidth]{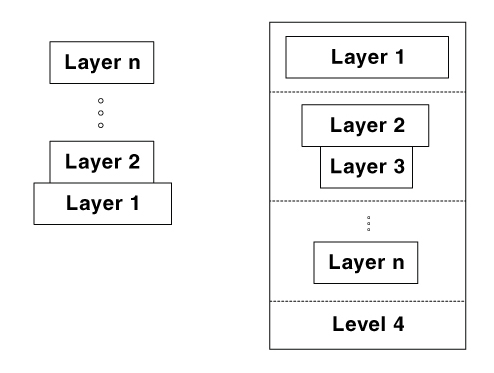}

\caption{Einstein's layers and the embedding to the classification scheme}
\end{figure}

It is surprising that within this layer model Einstein does not refer to an ontological level. Though the work is written in 1936, it relates to Einstein's early positions that are often seen in the positivistic tradition of Mach. A detailed account on Einstein's alleged shift from positivism to realism is given by \cite[]{howard1993einstein}.  \newline

\newpage

Einstein's model of layers is hierarchical and there is a maximum of logical unity a theory can develop to; it is represented by the highest layer in the model. The layer model is strongly influenced by the success in the evolution of Classical Mechanics from ancient conceptions to Newtonian Mechanics and General Theory of Relativity. It must have been this story of ongoing unification that motivated Einstein's view on theory structure. \newline
This view reflects also in the way Einstein presented the historical evolution of physics \cite[]{einstein1950evolution}.

\newpage

\subsection{The received view}

The received or syntactic view is the view on scientific theory that developed within the framework of Logical Empiricism in the 1920s in the environment of the Vienna Circle and influence of Machian positivism. A brief summary on the received view is provided by \cite[p.12]{suppe1977structure}:

\begin{quote}
\textit{``A scientific theory is to be axiomatized in mathematical logic (first-order predicate calculus with equality). The terms of the logical axiomatization are to be divided into three sorts: (1) logical and mathematical terms; (2) theoretical terms; and (3) observation terms which are given a phenomenal or observational interpretation.''}
\end{quote}

Between the theoretical terms and the observational terms, correspondence rules are defined. Every theoretical term has to have a corresponding observational term. If not, it would not be an allowed theoretical term within the received view; it would be what is seen as a metaphysical term, which has no relation to observations and according to the stance of Logical Empiricism can be neglected. \newline

These correspondence rules act between the Level 1 and 2 of the previously presented classification scheme. They do not refer to any metaphysical or ontological entity; they only represent relations between experience and theoretical terms.  

\begin{equation}
Tx \equiv Ox
\label{Correspondence rules in the received view}
\end{equation}

The received view failed due to several reasons, which we do not present in detail here. A detailed account is given by \cite[]{suppe2000understanding}, though two main reasons should be mentioned. Following \cite[p.103]{suppe2000understanding}, two critiques on the received view were essential:

\begin{quote}
\textit{``theories are not linguistic entities and thus theories are individuated incorrectly.''}
\end{quote}

and

\begin{quote}
\textit{``correspondence rules were a heterogeneous confusion of meaning relationships, experimental design, measurement, and causal relationships some of which are not properly parts of theories;''}
\end{quote}

\begin{figure}[h]
\centering

\includegraphics[width=0.5\textwidth]{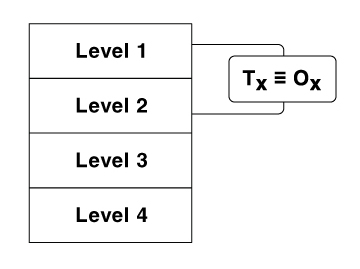}

\caption{The received view in the classification scheme}
\end{figure}

The received view faced criticism, specifically concerning the position on the meaning of theoretical terms. As pointed out by \cite[p.13]{suppe1977structure}, the summarized slogan of this position is

\begin{quote}
\textit{``The meaning of a term is its method of verification''.}
\end{quote}

Carl Hempel, a developer and main proponent of the received view, changed his position during the Illinois Symposium on the Structure of Scientific Theories in 1969 and questioned one of the cornerstones of the received view \cite[p.253]{hempel1977formulation}: 

\begin{quote}
\textit{``It presupposes that if theoretical terms have definite meanings, then it must be possible to construe those terms as introduced by specifiable logical procedures, which assign them meanings with the help of terms that are antecedently understood. \newline
This presupposition, which I once thought quite sound, seems to me mistaken for several reasons...''}
\end{quote}

This cleared the way for a new view on the structure of scientific theories, the semantic view, which developed out of the criticism on the received view. For an early presentation of the received view, refer to \cite[]{carnap1923aufgabe}; for detailed presentations of its failure, refer to \cite[]{hempel1970standard} and \cite[]{hempel1977formulation}.

\subsection{The semantic view}

The semantic view is a position that contrasted and succeeded the received view. It represents the view that a theory is represented by models, which makes use of mathematical terms and statements concerning theoretical terms.  

\cite[p.105]{suppe2000understanding}
\begin{quote}
\textit{``The Semantic Conception identifies theories with certain kinds of abstract theory-structures, such as configurated state spaces, standing in mapping relations to phenomena. Theory structures and phenomena are referents of linguistic theory-formulations. The basic idea is that theory structures are identified with suitably connected families of models. Depending on mapping relationships required for theoretical adequacy, realist, quasi-realist or antirealist versions are obtained.''}
\end{quote}

Whereas the received view neglected the need for metaphysical terms, the semantic view incorporates both possible polar positions of realism and antirealism on the question of ontological commitment \cite[p.106]{suppe2000understanding}: 

\begin{quote}
\textit{``Such commitments are via individual mapping functions (Loc functions) from real-world objects to points in logical space.''}
\end{quote}

and

\begin{quote}
\textit{``a realist has Loc functions onto every state variable and maintains a theory is empirically true just in case theory-structure-allowed state transitions are identical to those possibly occurring in the actual world. Antirealists do not commit ontologically to all state variables. They only require countenancing Loc functions from observables and that theories be empirically adequate: If W is that portion of reality to which one attaches Loc functions, the image M* of W is among the models comprising the theory.''}
\end{quote}

Within the semantic view, both positions can be represented, depending on which entities that the Loc functions refer to. Represented in the classification scheme presented before, the realist and antirealist positions refer to the ontological level (Level 4) or to observables inside the theories framework. 

\begin{minipage}{\textwidth}
\centering
\includegraphics[width=0.7\textwidth]{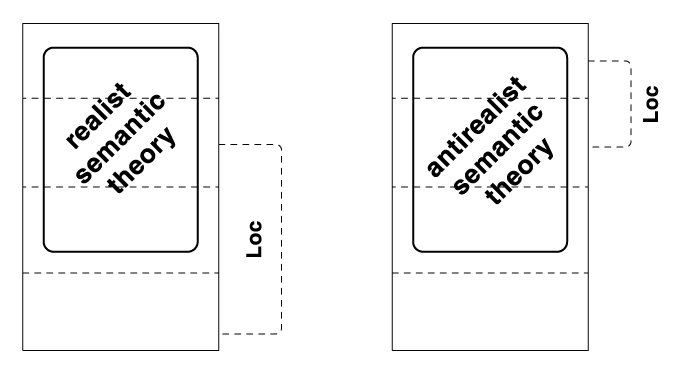}
\captionof{figure}{Realist and antirealist positions within the semantic view}

\end{minipage}
\bigskip

The semantic view itself has different versions, which have differences concerning the interpretation and role of confirmation \cite[]{daCosta1990model}, \cite[]{suppe2000understanding}, \cite[]{van1980scientific}.

Out of the semantic view, the structuralist approaches in philosophy of science emerged. 

\newpage

\subsection{The structuralist approaches}

There are several accounts on structuralism. Three main approaches are presented in \cite[]{sep-physics-structuralism}. We present them, in an overview and their relation to the classification scheme, in an overview.

An account on the development of the Structuralist Approach out of the Semantic approach is provided by \cite[p.108]{suppe2000understanding}:

\begin{quote}
\textit{``Like the Semantic Conception it analyzes theories set-theoretically as comprised of a theory structure and intended applications, but is neo-positivistic in spirit and reliance on a relativized theoretical/nontheoretical term distinction. It began with Sneed's (1971) application of Suppes's (1957) set-theoretic techniques to the problem of theoretical terms.''}
\end{quote}

\subsubsection{Sneed's structuralist approach}

The structuralist approach by Joseph Sneed is a continuation of the positivistic end of the semantic view. It utilizes the set-theoretic methods to model a scientific theory. Hand in hand with the application of set-theory goes the neglection of linguistic terms \cite[p.144]{sneed1976philosophical}:

\begin{quote}
\textit{``Roughly speaking the way of talking about scientific theories I am going to describe invites us to look at sets of "models" for these theories rather than the linguistic entities employed to characterize these models.''}
\end{quote}

Following the account in \cite[]{sep-physics-structuralism}, we present the conception of Sneed's approach. A recent introduction to these basic structures in Sneed's structuralism approach and their relation is provided by \cite[]{andreas2014basic}.

According to \cite[p.120]{sneed1976philosophical}, a scientific theory is a 

\begin{quote}
\textit{``... conceptual structure that can generate variety of empirical claims about a loosely specified, but not completely unspecified, range of applications.''}
\end{quote}

The structure is built by a core $K$ and a domain of intended application $I$. The core contains the class of potential models $M_p$, which is basically the set of possible models that describe the situation in terms of the used elements in the model. \newline

Assigned to the potential models, there is an actual model $M$ that represents the set of potential models that satisfy empirical laws. Together, they form a model-element $ \langle M_p,M \rangle $, the minimum element a theory has to contain.  \newline

$M_{pp}$ is the set of potential models without the theoretical terms that have been involved in $M_p$. 

As this $M_{pp}$ set is difficult to understand without a practical application, we refer to the example of Hookes law for massive point particles attached to a spring, presented in \cite[]{sep-physics-structuralism}. If $M_p$ represents the full conceptual frame with the functions of force, the spring constants, the masses of the point particles and the position functions of the particles, then $M_{pp}$ represents the model without all theoretical terms. This would reduce the partial potential models to the particles positions as mass, spring constant and forces as theoretical terms. \newline

Since the distinction between theoretical and non-theoretical terms is a peculiar question we refer to \cite[]{sep-theoretical-terms-science} for an overview of positions and criticism. \newline

Additional to $M_p$, $M$ and $M_{pp}$ the core contains constraints $C$ that connect different models in the same theory, a class of links $L$ that connect to models of other theories. For example, to describe the position of the point particles, an account on measuring distances in spacetime is needed \footnote{In the structuralist account of Ludwig presented in Section \ref{section Ludwig structuralist}, these links will be described by Pre-theories}. \newline

To give an account on the realistic situation of finite values in empirical sciences, a class of admissible blurs $A$ is introduced, which contains degrees of approximation \footnote{In the structuralist account of Ludwig presented in Section \ref{section Ludwig structuralist}, these blurs $A$ will be represented there by inaccuracy sets and unsharp measurements}. \newline

The theory core $K$ and the domain of intended application $I$ form a theory element $ \langle K,I \rangle $. The intended application $I$ refers to the empirical data that the theory refers to; in fact, to the phenomena \cite[p.125]{sneed1976philosophical}:

\begin{quote}
\textit{``The set $I$ is to be interpreted as the range of intended applications of the element - what the theory-element is about. The only requirement (D4) puts on $I$ is that its members have the structure characteristic of the non-theoretical part of $K$ - that they be members of $M_pp$.''}
\end{quote}

\begin{minipage}{\textwidth}
\centering
\includegraphics[width=0.4\textwidth]{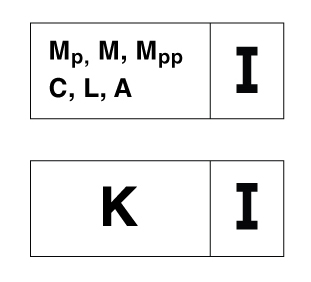}
\captionof{figure}{The theory core $K$ and the intended application $I$}

\end{minipage}
\bigskip

Theory elements $ \langle K,I \rangle $ form with other theory elements, a theory-net $N$. These theory-nets represent the complex we usually understand commonly as a scientific theory.

\begin{minipage}{\textwidth}
\centering
\includegraphics[width=0.9\textwidth]{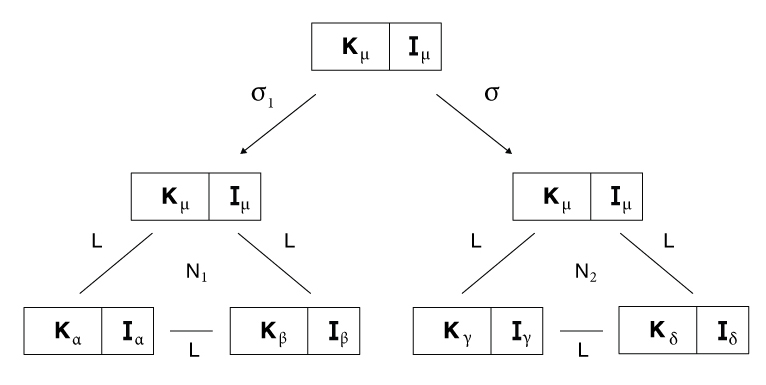}
\captionof{figure}{The different elements of the structuralist approach}

\end{minipage}
\bigskip

A theory-net that undergoes a change over time describes a theory evolution $I$ in time. 

One essential concept in this structuralist approach is the extension process, which describes the reasoning from a limited empirical dataset represented by T-nontheoretical terms to T-theoretical terms \cite[p.1459-1460]{gahde2014theory}:

\begin{quote}
\textit{``The starting point is provided by a certain set of data which is interpreted as a fragment of a model of an empirical theory. Attempts are made to extend this fragment into a complete model of that theory.
...
this extension process is to be described by referring to the distinction between theoretical and non-theoretical functions with respect to the theory T in question. The intended applications of that theory (which comprise the available data) are located at the non-theoretical level with respect to T and are represented by
so-called partial models. These partial models consist of base sets and T-nontheoretical functions. By adding suitable T-theoretical functions, they are to be
extended into models of T. ''}
\end{quote}

Concerning the question on scientific progress, Sneed provides an instrumentalist answer \cite[p.143]{sneed1976philosophical}:

\begin{quote}
\textit{``These considerations suggest that perhaps reduction of theories relative to specific nets is all we need to deal with the question of 'scientific progress'. They suggest that we do not need a concept of theory in which the possibilities for its future development are so strongly specified (through a strong specialization concept) that it can be established 'once and-for-all' that one theory can not keep up with another.''}
\end{quote}

Sneed's approach involves no reference to an ontology. It is a purely empirical account and is therefore confined to the Levels 1 and 2 of the presented classification scheme. Concepts and principles which are represented by Level 3 would contain linguistic terms and are neglected in the form of a statement by Sneed's structuralist approach. Principles like energy or momentum conservation are represented in the account, but not in the sense of a first order principle. They are merely empirical statements about and between the involved theory elements.

In contrast to Sneed's positivistic approach, a more realistic-oriented approach has been presented by G\"unter Ludwig.

\subsubsection{Ludwig's structuralist approach}
\label{section Ludwig structuralist}

G\"unter Ludwig presented an approach based on the axiomatization of Quantum Mechanics embedded in an axiomatization of scientific theories in general. An overview is provided by \cite[]{sep-physics-structuralism}:

\begin{quote}
\textit{``His(G. Ludwig's) underlying “philosophy” is the view that there are real structures in the world which are “pictured” or represented, in an approximate fashion, by mathematical structures, symbolically $ PT=W(-)MT $. The mathematical theory $MT$ used in a physical theory PT contains as its core a “species of structure” Σ. This is a meta-mathematical concept of Bourbaki which Ludwig introduced into the structuralistic approach. The contact between MT to some “domain of reality” $W$ is achieved by a set of correspondence principles  $(-)$, which give rules for translating physical facts into certain mathematical statements called “observational reports”.''}
\end{quote}

\cite[p.3]{ludwig2007new}
\begin{quote}
\textit{``The reality is in part constituted of facts stating "basic properties" of objects and "basic relations" between objects. Only facts related to the "domain of physics" are taken into consideration. These facts, directly recordable or indirectly recordable via known theories, called pre-theories, constitute what we call the physically recordable domain or the reality domain.''}
\end{quote}

Ludwig introduces the basic structure of a physical theory as the statement $PT=W(-)MT$. It is a correspondence between the reality domain and a mathematical theory. In his last account \cite{ludwig2007new} altered the notion of the correspondence principle and replaced the usage to the symbol $(-)$, as a representation of the correspondence between the reality domain and the mathematical theory, with the symbol $(cor)$. 

What Ludwig defines as an application domain is introduced by Einstein in his model of layers as "first layer" \cite[p.12]{ludwig2007new}:

\begin{quote}
\textit{``The application domain of a particular physical theory $PT_\nu$, denoted by $A_{p\nu}$ is the restriction of the reality domain $W_\nu$ to the facts that the theory considers a priori.
The recording of facts can be made directly or indirectly by pre-theories.''}
\end{quote}

A pre-theory is a method to record those a priori given facts; it could, for example, be a pre-theory of distance measurement. The application domain is the collection of all those facts that build a basis for the theory $PT_\nu$. The application domain of a theory can include reality domains $W_\alpha$ $W_\beta$ of other physical theories $PT_\alpha$, $PT_\beta$; they serve as pre-theories to the physical theory $PT_\nu$.

The account of Ludwig contains one specific feature, the consideration of the finiteness of physics. He introduces inaccuracy sets to give regard to the fact that numbers of observations are finite, and the mathematical axiomatization is an idealization which involves an unavoidable inaccuracy. 

This also reflects in the domains of physical reality. Additional to the application domain of a theory, he introduces a fundamental domain $G_\nu$ \cite[p.13]{ludwig2007new}: 

\begin{quote}
\textit{``The fundamental domain of a particular physical theory $PT_\nu$, denoted by $G_\nu$ is the restriction of the application domain $A_{p\nu}$ to the facts that the theory describes.
...
It is often more useful to apply the theory only on that part of the application domain $A_{p\nu}$ where we can use a small degree of inaccuracy. In such a region the theory essentially says something about the structure of reality and will be useful for technical application.''}
\end{quote}

If the theory $PT_\nu$ provides small inaccuracies over the whole application domain, the application and foundational domain are equivalent $G_\nu \equiv A_{p\nu}$.

At last there is a reality domain $W_\nu$ of the physical theory $PT_\nu$ \cite[p.14]{ludwig2007new}:

\begin{quote}
\textit{``The reality domain of a particular physical theory $PT_\nu$, denoted by $W_\nu$, is the extension of the fundamental domain $G_\nu$ to the facts(related to the new physical concepts) that the theory describes.
Our task is not only to detect "nonmeasured" realities, but also to detect new realities.''}
\end{quote}

The connection between the three domains of physical reality and the mathematical theory $MT$ is shown in Figure \ref{ludwigrealitydomains}.

\begin{minipage}{\textwidth}

\centering
\includegraphics[width=0.3\textwidth]{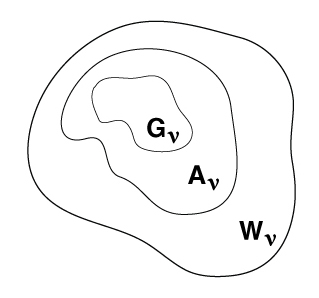}
\captionof{figure}{The three domains of physical reality according to Ludwig}
\label{ludwigrealitydomains}
\end{minipage}
\bigskip

The three domains of physical reality contain each other as a subset. 

\begin{equation}
A_{p\nu} \subset G_\nu \subset W_\nu 
\end{equation}

\begin{minipage}{\textwidth}
\centering
\includegraphics[width=0.25\textwidth]{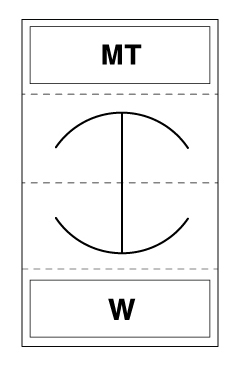}
\captionof{figure}{The connection between Ludwig\textsc{\char13}s structuralism approach and the classification scheme}
\label{ludwigschemecomparision}
\end{minipage}
\bigskip

The reality domain $W_\nu$ is corresponding, to our definition of the ontological Level 4 of a physical theory. The mathematical theory corresponds to the Level 1 and the correspondence rules $(cor)$ between the mathematical theory and the domain of physical reality, which contain the interpretations and the concepts, refer to the Levels 2 and 3 of the classification scheme. \newline

Ludwig's approach is at the realistic end of the spectrum of structuralist accounts. 

\newpage

\subsubsection{Scheibe's structuralist reductionism approach}

The third structuralist approach is Erhard Scheibe\textsc{\char13}s account and refers mainly to the mechanisms of reduction of theories. According to \cite[]{sep-physics-structuralism}, his view can be seen between Sneed\textsc{\char13}s and Ludwig\textsc{\char13}s approaches. Although Scheibe\textsc{\char13}s approach is not relevant to our discussion on a classification scheme, we mention his approach and its relation in the context of structuralist accounts. For an overview, refer to \cite[]{scheibe2001between} and \cite[]{scheibe2013reduktion}. \newline

To summarize the idea of structuralist approaches, we refer to \cite[p.399]{balzer1980gunther}:

\begin{quote}
\textit{``Die These des Strukturalismus l\"a\ss t sich kurz so fassen: Erkenntnis der Wirklichkeit besteht darin, aus ihr Strukturen herauszulesen und in sie Strukturen
hineinzuinterpretieren.''}
\end{quote}

\begin{quote}
\textit{``The approach of structuralism explained in short: Insight about reality persits in perceiving the structures out of nature and interpreting structures into the realm of reality.\footnote{Translation by the author}''}
\end{quote}

\subsection{Epistemological and ontic structural realism}

Out of the structuralist ideas, new approaches evolved. Structural realism emphasizes the importance of the structural relations and assigns to them central relevance. The ontological role of objects in structural realism is a matter of dispute, and the different positions within structural realism assign different weight to the concept of reality of objects \cite[]{sep-structural-realism}:

\begin{quote}
\textit{``The structuralist solution ... is to give up the attempt to learn about the nature of unobservable entities from science. The metaphysical import of successful scientific theories consists in their giving correct descriptions of the structure of the world. Theories can be very different and yet share all kinds of structure. The task of providing an adequate theory of approximate truth that fits the history of science and directly addresses the problem of ontological continuity has hitherto defeated realists, but a much more tractable problem is to display the structural commonalities between different theories.''}
\end{quote}

This abandoning of unobservable entities has different degrees. Within the structural realism, the bandwidth goes from a a realistic-oriented epistemological structural realism (ESR) to a positivistic-oriented position of ontic structural realism (OSR).   

\newpage

\subsubsection{Epistemological structural realism (ESR) }

In the framework of the discussion of realist positions and anti-realist positions on the meaning of scientific progress and the value of ontological statements, three main arguments are noteable: The miracle argument by \cite[]{putnam1975mathematics} defending the position of realism, the Scientific Underdetermination \cite[]{sep-scientific-underdetermination} and the Pessimistic Meta Induction by \cite{laudan1981confutation} on the side of anti-realism. \newline

John Worrall attempted to find a synthesis of these ideas and asked if it is possible to get the best of both worlds \cite[p.101]{worrall1989structural}:

\begin{quote}
\textit{``The main interest in the problem of scientific realism lies, I think, in the fact that these two persuasive arguments appear to pull in opposite directions; one seems to speak for realism and the other against it; yet a really satisfactory position would need to have both arguments on its side. The concern of the present paper is to investigate this tension between the two arguments and to suggest (no more)
that an old and hitherto mostly neglected position may offer the best hope of
reconciling the two.''}
\end{quote}

The way that the ESR attempts to resolve the miracle argument is basically the statement that what is common to an outdated theory and a successor theory are the relations between elements of the theories, the structures. In the outdated theory, they described reality accurately enough, at least in a specific regime, so that the old description was eligible. The successor theory extended the application domain and incorporated the outdated old theory. The ontological commitments might have changed while proceeding to the new theory; the old theory might have introduced ontological entities that are superfluous in the successor theory, but the structures between objects are common in both theories. Therefore, ESR gives attention to these structures, outlined in an example on the theory of light \cite[p.117]{worrall1989structural}:

\begin{quote}
\textit{``There was an important element of continuity in the shift from Fresnel to Maxwell - and this was much more than a simple question of carrying over the successful empirical content into the new theory.
... There was continuity or accumulation in the shift, but the continuity is one of form or structure, not of content.''}
\end{quote}

Concerning metaphysical concepts, ESR takes the position that metaphysical statements have to follow our best theories and are not to be seen as first-order principles \cite[p.123]{worrall1989structural}:

\begin{quote}
\textit{``The only claim is that ultimately evidence leads the way: if, despite all efforts, no scientific theory can be constructed which incorporates our favourite metaphysical assumptions, then no matter how firmly entrenched those principles might be, and no matter how fruitful they may have proved in the past, they must ultimately be given up.''}
\end{quote}

Emphasizing the structures as the elements of a theory that ensure continuity in the shift from an outdated theory to a successor theory does not mean that ESR neglects the objects or entities on the ontological level, as those elements are connected by structures. Concerning mappings from theory elements to the ontological level, ESR advances a position considerably different than structuralist approaches before \cite[p.154]{worrall2007miracles}: 

\begin{quote}
\textit{``But there is no reason why the way in which a theory mirrors reality should be the usual term-by-term mapping described by
traditional semantics.''}
\end{quote}

In the context of the classification scheme, ESR could be seen as giving credit to a reality for objects, but not in an accessible way. The structures correspond to elements we can access, but there is more on the ontological level, the relata connected by the structures. These elements referring to the objects are epistemologically hidden to us.

\begin{minipage}{\textwidth}
\centering
\includegraphics[width=0.5\textwidth]{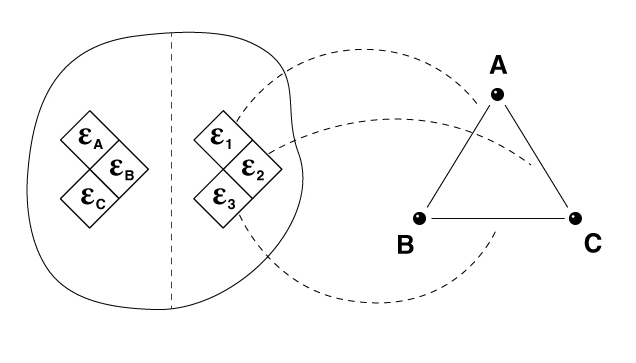}
\captionof{figure}{ESR in the classification scheme}
\end{minipage}
\bigskip

And further concerning this mapping \cite[p.154]{worrall2007miracles}:

\begin{quote}
\textit{``SSR\footnote{Structural Scientific Realism, is the wording John Worral assigned to this program by himself. Ladyman introduced ESR} in fact takes it that the mathematical structure of a theory may globally reflect reality without each of its components necessarily referring to a separate item of that reality.''}
\end{quote}

This aspect involves a specific form of holism in the conception of ESR that we cannot discuss in further detail here, but the relation to other holistic conceptions such as \cite[]{bohm1987ontological}, \cite[]{quine1951main} and for a recent review on confirmational holism \cite[]{carlson2015logic}, is an open question. \newline 

ESR holds a scientific realism that denies that scientific theories provide access at all to these elements of reality \cite{worrall2007miracles}:

\begin{quote}
\textit{``But why should a realist not be equally as fallibilist and tentative about the mode of reference of the terms in theories as she is about those theories truth? In any event, no one seriously holds, as I have remarked several times, that we have any theory-independent access to the furniture of the universe that would allow us to
compare (even ‘in principle’) the notions conjectured by our theories with what there really is.''}
\end{quote}

To proponents of ESR, this inaccessibility is not an argument in favour of anti-realism. In contrast to ESR, Ontic structural realism has been introduced as a positivistic point of view on the structural realist account.

\subsubsection{Ontic structural realism (OSR) }

The origin of the ontic structural realism (OSR) is the distinction between epistemological and ontic structuralist approaches provided by \cite[]{Ladyman1998409}. OSR as well as ESR emphasizes the role of structures as central elements of ontology \cite[]{sep-structural-realism}: 

\begin{quote}
\textit{``Ontic structural realists argue that what we have learned from contemporary physics is that the nature of space, time and matter are not compatible with standard metaphysical views about the ontological relationship between individuals, intrinsic properties and relations. On the broadest construal OSR is any form of structural realism based on an ontological or metaphysical thesis that inflates the ontological priority of structure and relations.''}
\end{quote}

The essential novelty of OSR is the neglection of the individual objects and their properties as metaphysical and superfluous. Therefore, Ladyman´s position is often referred to as radical OSR. A thorough review of the manifold views and peculiarities concerning the difference between epistemological and ontic structural realism is presented in \cite[]{sep-structural-realism}. \newline

It is notable that OSR is a realist account, but confined to the structures. If we identify the empirical content of the positivistic approaches by Sneed with the structures in OSR, OSR could be seen in the positivistic tradition of Logical Empircism. Then, OSR is a positivistic account concerning properties of objects and individuality. \newline 

In the classification scheme, the OSR approach would refer to the ontological Level 4, but only regarding the structures. Therefore, the identification of elements of reality is not due to properties of objects; it refers to structures and relations.  

\begin{minipage}{\textwidth}
\centering
\includegraphics[width=0.4\textwidth]{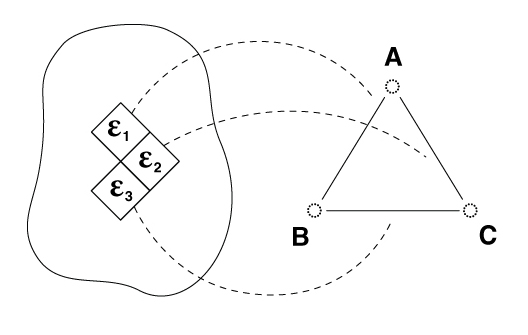}
\captionof{figure}{OSR in the classification scheme}
\end{minipage}
\bigskip

One account that could be thought of as intermediate between ESR and radical OSR is advocated by \cite[]{esfeld2008moderate}. Moderate OSR holds the position that relations without relata are inconceivable. Challenging the radical version, \cite[p.29]{esfeld2008moderate} states:

\begin{quote}
\textit{``If it is claimed that there is something that exists but that we cannot know, we need an argument why we should accept that there is any such thing. The master argument for intrinsic properties can be summed up in this way:\newline
(1) Relations require relata, that is, objects that stand in the relations. \newline
(2) These objects have to be something in themselves, that is, they necessarily have
some intrinsic properties over and above the relations that they bear to one
another—even if the relations do not supervene on the intrinsic properties and
even if we cannot know the intrinsic properties.''}
\end{quote}

Moderate OSR takes a different position concerning the ontological status of the objects and the relation. Following moderate OSR, they are ontologically on the same level; none is primary. 

\begin{minipage}{\textwidth}
\centering
\includegraphics[width=0.4\textwidth]{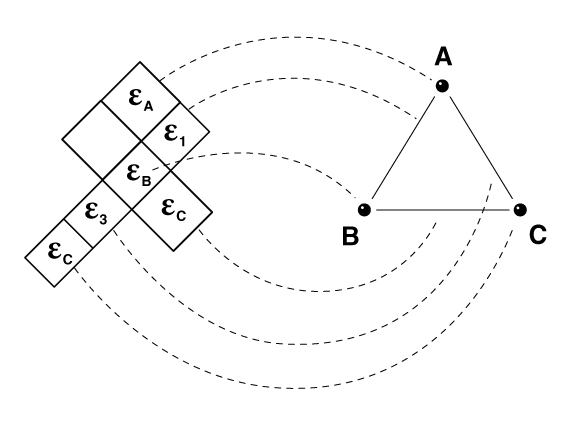}
\captionof{figure}{Moderate OSR in the classification scheme}
\end{minipage}
\bigskip

In the context of the presented classification scheme, this implies that there exists elements of reality for the properties and the structures on the same footing.

\begin{minipage}{\textwidth}
\centering
\includegraphics[width=0.8\textwidth]{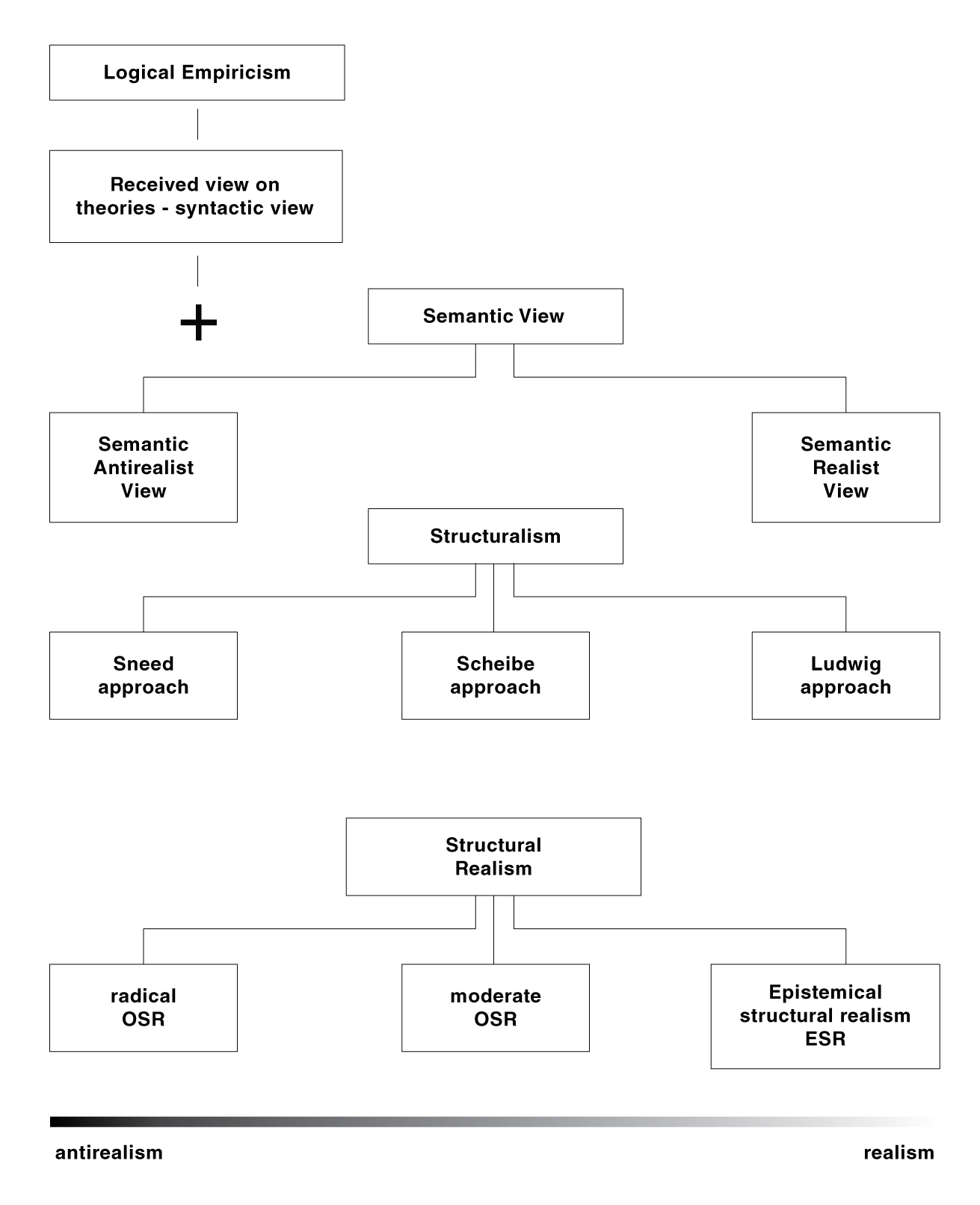}
\captionof{figure}{An overview on the discussed points of view with respect to realism and antirealism}
\end{minipage}
\bigskip

We want to close this presentation of accounts on the structure of scientific theories with an overview of the previously presented approaches and their relation to each other, and the dichotomy of positivism and realism.

\newpage 

\subsection{Quine-Duhem thesis}

The Quine-Duhem thesis is an approach that understands science as a system of statements that are connected with each other. There are statements that are far from experience; they form the center of a imagined sphere. Those statements involve logic and mathematics. Located on the outer regions of the sphere, connected to the realm of mathematics and logic, are those statements that are in connection with experience. The scientific enterprise is represented by the structure of this web of belief. \newline

\begin{minipage}{\textwidth}
\centering
\includegraphics[width=0.6\textwidth]{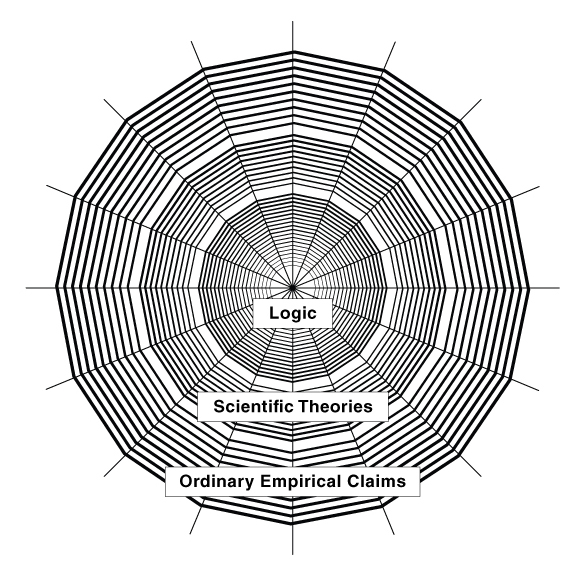}
\captionof{figure}{The web of belief}
\end{minipage}
\bigskip

If science and the network of statements that forms our web of belief gets in contradiction with experience, the whole web of belief is in contradiction. The Quine-Duhem thesis will state the following: \newline

It is equivalent, in our intention, to remove the contradiction if we add or modify a single statement on the outer shores of the web of belief, or if we modify statements at the outermost heart of our system and thereby completely change the whole web of belief. \newline

The conception is called confirmational holism; it expresses that we never perform empirical tests in isolation. It challenges falsification, according to Popper as well, because falsification would refer also to the whole web of belief. In the sense of a parsimony principle of necessary modification during a contradiction, it is worthwhile to aim for a minimum of needed modifications. This stance is called Conservatism and is connected to the social psychology aspect of the scientific enterprise \cite[Chapter 2.1]{sep-scientific-underdetermination}; keep as much as possible, introduce as much as necessary.  

A recent example for the application of the Quine-Duhem thesis is the interpretation of Quantum Mechanics. Although there is no contradiction with experience, the interpretations of Quantum Mechanics demonstrate a situation where different sets of statements at the shores of the web of belief seem to be equivalent. The Quine-Duhem thesis would see those interpretations as equivalent, and there are even interpretations that demonstrate the change at the very heart of the system. Quantum Logic is an approach in the interpretation of Quantum Mechanics that modifies the statements in the realm of logic. For an introduction, refer to \cite[]{mittelstaedt1978quantum} or \cite[]{van1980scientific}.

\section{Statements about the ontological level}

Statements that refer to metaphysical entities refer to the Level 4 of the classification scheme presented previously. Logical Empiricism denied the existence of statements of that kind. We present our views on possible statements on the metaphysical level.

\subsection{Ontological coherence}

The ontological coherence argument (OCA) is a heuristic that supports the process of theory development from the perspective of ontological consistency in a network of physical theories. Those theories offering ontological elements that fit coherently with ontological elements from a different regime are preferable. By the mutual consolidation of the ontological implications, the OCA supports a network of theories that provides coherent ontologies. Thereby, OCA strongly supports scientific realism and related ideas.

\begin{theorem}
Under the assumption of a set of Theories $T_{PR \mu}$, which covers different regimes of physics, and that $T_{PR \mu}$ are theories in agreement with empirical data in the respective regime, the OCA claims that those theories $ T_{PR\nu}\in T_{PR \mu}$, whose ontologies fit coherently together, are preferable. 
\end{theorem}

We define a set of elements of a theory $ T_{PR} $ and a set of elements of reality $ PR $:

\begin{equation}
T_{PR\nu} = \{ t_{\nu i} \} = \{ A_\nu,B_\nu,C_\nu,... \}
\label{Definition Elements of theory}
\end{equation}

\begin{equation}
PR_\nu = \{\epsilon_{\nu i}\} = \{ \epsilon_{\nu A},\epsilon_{\nu B}, \epsilon_{\nu C},... \} 
\end{equation}

The set of elements of the theory and the elements of reality obey correspondence rules that are part of the theory. We do not put detailed emphasis on these correspondence rules. A detailed account on the mathematical definition of these rules is provided by \cite[]{ludwig2007new}. \newline

We call two theories $ T_{PR} $ theory-coherent if their corresponding elements of reality fulfill the condition

\begin{equation}
PR_\alpha \cap PR_\beta \neq \emptyset.
\end{equation}

This implies non-empty sets of elements of reality shared by two theories. In the framework of scientific realism and its related ideas, this provides an argument for an ontological value of those elements of reality.

\begin{minipage}{\textwidth}
\centering
\includegraphics[width=0.5\textwidth]{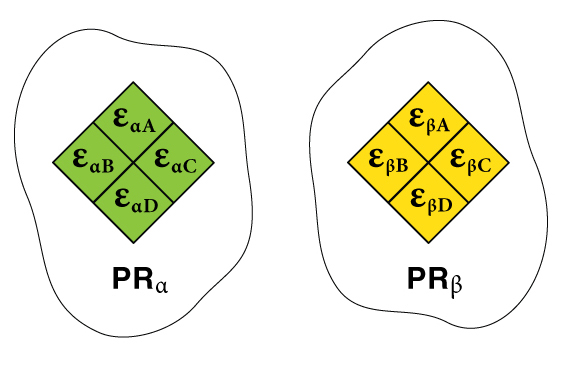}
\captionof{figure}{The sets of elements of reality of two theories}
\end{minipage}
\bigskip

\begin{minipage}{\textwidth}
\centering
\includegraphics[width=0.4\textwidth]{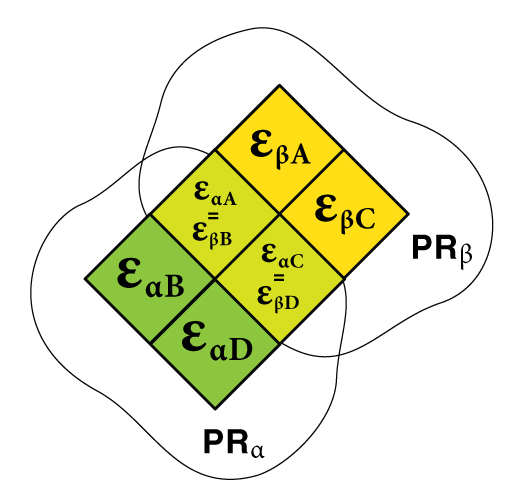}
\captionof{figure}{Two sets of elements of reality of two theories}
\end{minipage}
\bigskip

The idea of OCA is related to Ludwig's account on the structure of scientific theories. He introduced the reality domain of a physical theory and stated on the constructed reality domain of all physical theories \cite[p.14]{ludwig2007new}:

\begin{quote}
\textit{``We can now add that the reality domain $W$ is the domain of all $W$s, i.e., the $W$s of all $PT$s. Given that all $PT$s are not known, $W$ cannot be established. By finding new $PT$s, one discovers new $W$s(e.g., atoms and elementary particles). The physically recordable domain $W$ remains decisively limited by the fact that one does not permit all directly ascertainable facts such as, e.g., that a sound is harmonious or that a violin has a good sound.''}
\end{quote}

And in an earlier account \cite[p.186]{ludwig1978grundstrukturen}:

\begin{quote}
\textit{``Durch die Weiterentwicklung der Physik wird der durch alle $PT$´s zusammen erfassbare Teil der Wirklichkeit immer größer. Beim Aufstellen des Wirklichkeitsbereichs $W$ einer $PT$ kann bei diesem Entwicklungsprozess des Erfassens eines Ausschnittes der Wirklichkeit kein Widerspruch auftreten...; dagegen sind beim Zusammenfügen der Wirklichkeitsbereiche der verschiedenen $PT$´s im Laufe der historischen Entwicklung der Physik immer wieder Widersprüche der einzelnen Wirklichkeitsbereiche untereinander und der Aussagen der Theorie mit der Erfahrung entstanden.''}
\end{quote}

\begin{quote}
\textit{``By the evolution of physics the determined part of reality is growing. In the definition of a reality domain $W$ of a physical theory $PT$ no contradiction can arise in this process of recognition of a part of reality...; In contrast in the history of physics during the merging of reality domains of different $PT$´s  contradictions arose at all time within reality domains and between statements of theories and empirical evidences.\footnote{Translation by the author}''}
\end{quote}

One specific contradiction in the reality domains that Ludwig refers to is the contradiction between the particle picture and the wave picture in Quantum Mechanics. We will give an account of this contradiction in Section \ref{Section Ontological truth}, after the definition of ontological truth.

\newpage

\subsection{Ontological truth}
\label{Section Ontological truth}

Logical Empiricism avoids to provide ontological statements. In our context of the classification scheme, this alludes to statements that refer to the Level 4. Statements of that kind enclose metaphysical content and are superfluous in the sense of Logical Empiricism's doctrine to confine itself only to statements on experience. If we ask for truth values of statements on metaphysical content, Logical Empiricism would deny that such statements are meaningful. \newline

Critical Rationalism would also deny that truth values on metaphysical statements are meaningful at all. In Critical Rationalism, the argument would be that metaphysical statements cannot be falsified by experience if the metaphysical content does not refer to any experience \cite[]{popper2005logic}. \newline

We claim that this does not imply that scientific knowledge is not growing, or that we can not learn more about how the world "really" is. Assuming that it is possible to learn about such a reality, which exists independent of us, is the basic assumption of scientific realism and all its related positions. The assumption is that there are statements on the metaphysical level that have truth values. \newline

Objections against such a view are manifold. Most important to the author are Scientific Underdetermination of theories\cite[]{sep-scientific-underdetermination} and the Pessimistic Meta Induction (PMI)\cite[]{laudan1981confutation}. A summary of the PMI argument is provided by \cite[]{sep-structural-realism}: 

\begin{quote}
\textit{``Proposition p is widely believed by most contemporary experts, but p is like many other hypotheses that were widely believed by experts in the past and are disbelieved by most contemporary experts. We have as much reason to expect p to befall their fate as not, therefore we should at least suspend judgement about p if not actively disbelieve it.''}
\end{quote}

For recents debates on the PMI argument, see \cite[]{fahrbach2009pessimistic}; \cite[]{park2011confutation} and  \cite[]{muller2015pessimistic}. \newline

In favour of scientific realism, Hilary Putnam formulated the Miracle argument \cite[]{putnam1975mathematics}, which basically states that the success of explanation and prediction of our best scientific theories would be a miracle if these theories do not refer to a given reality beneath the level of theories. \newline 

Mathematical realists would see here a way with their approach of ontological reality of mathematical entities to solve the "miracle" in the miracle argument, the unreasonable effectiveness of mathematics \cite[]{wigner1960unreasonable} and the question on the nature of this metaphysical reality in one stroke. \newline

The question on the nature of reality is still open, and as the recent discussions on structural realism demonstrate, an end is not in sight. Too many questions are open or unclear and need a thorough analysis, but we want to present a simple argument that provides a preliminary and confined answer to the question of scientific knowledge and scientific progress:

\begin{theorem}
There exist statements on truth values on the level of ontologies that result from empirical science. These statements always are exclusive statements on ontologies.
\end{theorem}

Example: The earth is not a disk. \newline

Example: The earth is not a sphere. \newline

These statements are both true and can claim to refer to the ontological level. We can state that the earth is not a disk, since we have an idea about the theoretical term "disk" and its ontological meaning. From experience, we could deduce that the earth is not a disk. And the same for the statement that the earth is not a sphere. The earth has an eccentricity and has properties that definitely do not resemble those of a sphere. That is a statement that follows from experience and refers to the ontological level of the entity earth. \newline

We cannot provide an answer as to what the earth or any ontological entity is, but we have access to empirical statements that help us in our quest on scientific knowledge. One could give the objection that this is not very satisfying, because the realm of those metaphysical entities which the earth potentially resembles is very large, probably infinitely large. The statements we can provide only are finite and it gives the impression that although we gain scientific knowledge, we will not come to an end. 

\begin{minipage}{\textwidth}
\centering
\includegraphics[width=0.4\textwidth]{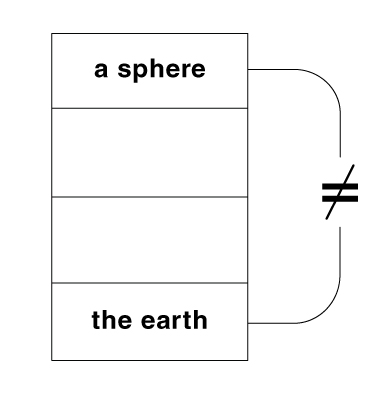}
\captionof{figure}{The relation of the terms to the ontological levels in the statement}
\end{minipage}
\bigskip

Another debate that demonstrates the application of the concept of ontological truth is the discussion on the Wave-particle duality, or in Bohr's conception Wave-particle complementarity. An introduction to the historical development of methodologies of particle and wave-theory of light is provided by \cite[]{achinstein1991particles}. An overview on the positions of wave-particle complementarity concerning light is provided by \cite[]{rauchcombourieu1992wave} and \cite[]{ghose1996two}. \newline

The classical concepts of particles and waves are, in principle, untenable in a general way. A particle that is localized at a specific point in spacetime, carrying properties, is outworn in the same way as the concept of a completely delocalized and holistic entity wave that is spread over space and time, reaching into infinity. Both concepts are based on the Level 3 of the classification scheme and represent an idea on an ontological entity. Following the conception of ontological truth, a statement of the kind "light is a particle" or "light is a wave" cannot be a tenable statement on truth about the ontological entity light. Also, a statement of the kind "light is both particle and wave" is not valid according to the conception of ontological truth. \newline

The negated statements "light is not a particle" and "light is not a wave" would fulfill the requirements on truth statements, as well as "light is neither a particle nor a wave." It turned out that the classical conceptions particle and wave do not represent concepts that fulfill the needs of our best theories. \cite[p.1403-p.1404]{rauchcombourieu1992wave} provide a review of why these conceptions failed, and Bohr had to introduce a principle to recover the classical conceptions:

\begin{quote}
\textit{``... no experimentalist has ever been able to show experimentally this "dual" behavior within the same experiment with a single particle and an
experimental setup absolutely identical; in other words, to violate the complementarity principle postulated by Niels Bohr in 1927, and since then accepted by the majority of physicists as an operational principle to settle the wave-particle dualism problem.
...
No experimentalist ever has succeeded in detecting a single particle travelling in one or the other branch of an interferometer (Michelson, Mach-Zehnder, or pure crystal) while analyzing simultaneously the interference pattern which it is supposed to appear on a screen located just behind the detectors''}
\end{quote}

Out of our best empirical theories, we get an idea about the properties and behaviour of light; we can model a conception, but never unveil the whole truth about an ontological entity. 

\cite[]{callender2015one} provides an account on resolving the wave-particle duality by embedding the question into a greater philosophical framework and assuming only one type of ontological entity:

\begin{quote}
\textit{``We have motivated the nomological understanding of the wavefunction. On this picture the It-or-Bit debate, and even the wave-or-particle debate, are absorbed by a larger more philosophical debate over the nature of laws.
...Regardless of whether one explores this option, the main point is
that quantum mechanics requires only one type of beable.''}
\end{quote}

The relation to scientific underdetermination and the unreasonable effectiveness of mathematics should be discussed within the framework of ontological truth and remains an open question. \newline

There is a manifold of philosophical approaches on the question of truth values of scientific statements. One prominent account is provided by \cite{tarski1944semantic}. All those approaches are philosophical and exist somewhat parallel to each other and it seems that they cannot be falsified by empirical evidences. The goal of the stated theorem on ontological truth is to provide an insight on progress in the quest for scientific knowledge, even if that implies to deny or stay agnostic towards a convergent realism. Further, it clarifies that there are statements with truth values even on the ontological level. That these statements have to be formulated in a negative exclusive way resembles the falsificationism in the empirical sciences.

\newpage

\section{Application to Physics and the Interpretations of Quantum Mechanics}

\subsection{Classical particle mechanics}

If we apply the classification scheme presented previously to specific physical theories and the interpretations of Quantum Mechanics, it is reasonable to demonstrate these principles by means of a well-known physical theory. We have chosen classical particle mechanics to demonstrate the application of the classification scheme. \newline

A detailed presentation of classical particle mechanics by means of Sneed's structuralist approach has been provided by \cite[]{ishigaki1995formal}. We will demonstrate the application of the classification scheme in a brief, incomplete, but illustrative way.

\subsubsection*{Level 1 - The mathematical formalism}

The mathematical formalism is set up by vector valued functions $\vec  x$ and $\vec  F $ in the vector space $\mathbb{R}^3$ and a scalar $m$. The vectors or rather their components are in generality functions of the free parameter time. The vector $\vec  x(t)$ has a derivation $\frac{d\vec x}{dt}$ we denote with $\vec v$. \newline

\begin{equation}
\frac{d(m\vec v)}{dt}=\vec F
\end{equation}

This is a linear second order differential equation. A solution is given by $\vec  x(t)$.

\subsubsection*{Level 2 - Physical quantities and connection to empirical rules}

In the Level 2 of the classification scheme, the physical quantities and their representation by the mathematical symbols introduced in Level 1 are defined.

\begin{mydef}
The physical quantity "position" is represented by the vector $\vec x$.
\end{mydef}
\begin{mydef}
The physical quantity "time" is represented by the parameter $t$.
\end{mydef}
\begin{mydef}
The physical quantity "mass" is represented by the scalar $m$.
\end{mydef}
\begin{mydef}
The physical quantity "force" is represented by the vector $\vec F$.
\end{mydef}

The physical quantity "position" represents results of measurements of the position coordinate of an object; for an extensive object, this will be the center of mass coordinates, in case of a pointlike object, it will be the coordinates of the point. 

We adopt the SI-unit system to relate these physical quantities with empirical statements. 

One peculiarity is that we cannot define a measurement procedure for the physical quantity force independently of the acceleration of the object unless we introduced a source of a force, such as Newton's gravitational law, which we will omit. 

\subsubsection*{Level 3 - Concepts and principles}

We define the concept of a "point particle" to describe the behaviour of extensive objects and pointlike entities as well: 

\begin{mydef}
A point particle is defined by its position coordinate in space represented by a vector $\vec x$ in the vector space $\mathbb{R}^3$, the parameter time $t$ as well as a set of associated properties. In the case of classical particle mechanics, this set is containing the physical quantity "mass."
\end{mydef}

The full power of the formalism of classical particle mechanics is available when we relate it with other specific laws of force. Therefore, the conception of a force is needed.

\begin{mydef}
A force is an interaction between an object and the origin of the specific force. The object will change the motion of the object according to the defined law, depending on the quantity of the force and its set of properties.
\end{mydef}

With that definition of force, concrete laws of force can be introduced, such as the gravitational law, Hooke's law, or friction. This interrelates the theory to other theories, which is the idea of the conception of links in Sneed's structuralist approach. 

\subsubsection*{Level 4 - Ontology}

The ontological statements on classical particle mechanics involve connected ontologies of space and time, according to the OCA, and involves them in the ontological level of the classical particle mechanics. \newline

Classical particle mechanics is based on a particle and matter ontology in which there are entities named particles made of matter that exist in a pre-existing space and alter their positions due to a time parameter according to real influences and interactions by forces. The forces are on the same ontological footing, in that sense as real, as the particles and represent the interaction process between particles with other entities. This ontology is deeply related to materialism, which has the view that what there is, is a matter in motion in a pre-existing space and time framework.   

\newpage
\subsubsection*{Overview}

\begin{table}[h]
\centering

\label{SchemeNewton}
\begin{tabular}{|l|c|}
\hline
\begin{tabular}[c]{@{}c@{}}\\ 1 \\ \\ \end{tabular}  &  $m{\ddot  {{\vec  x(t)}}}=\sum\limits_{i=1}^N{{\vec  F}(t)_{i}} $  \\  \hline
\begin{tabular}[c]{@{}c@{}}\\ 2 \\ \\ \end{tabular} & \begin{tabular}[c]{@{}c@{}}  \\ x... center-of-mass coordinate of an object  \\ m ... interial mass of an object\\ $\sum\limits_{i=1}^N{{\vec  F}_{i}} $ ... sum of all forces\\ \\ Measurement rules for $\vec F, m, \vec  x(t), t  $  \\  \\ \end{tabular} \\ \hline
\begin{tabular}[c]{@{}c@{}}\\ 3 \\ \\ \end{tabular} & \begin{tabular}[c]{@{}c@{}} \\ Point particle concept \\  \\ \end{tabular}                                                                                 \\ \hline
4 & \begin{tabular}[c]{@{}c@{}}\\ Entity of matter - Materialism \\ Position in space and time \\ \\ \end{tabular}                      \\ \hline
\end{tabular}
\caption{Classical particle mechanics}
\end{table}

\newpage

\subsection{De Broglie-Bohm interpretation}
The De Broglie-Bohm interpretation is a realistic, deterministic, nonlocal\footnote{Locality is the concept that events are independent of all other events that are space-like separated. Nonlocality abolishes this restriction, and therefore events can have an instantaneous influence on space-like separated events, but superluminal signalling is still restricted in this interpretation.} hidden-variable interpretation and is based on the decomposition of the Schr\"odinger equation into a set of two equations. One equation is interpreted as the conservation of probability, and the other equation is interpreted as a Hamilton Jacobi equation with an extra term. This term is interpreted as a Quantum Potential. From the Hamilton Jacobi equation, an equation of motion for the particles can be derived, which is called guidance equation. In the De-Broglie Bohm interpretation, nonlocality is an important and inherent feature of Quantum Mechanics, but is only an aspect of a much deeper principle that is central to the De Broglie-Bohm interpretation, the concept of ontological holism. The De Broglie-Bohm interpretation must be distinguished from Bohmian Mechanics, which is based on the same formalism, but adheres to a different conception and ontology. We will present Bohmian Mechanics in section \ref{section Bohmian Mechanics}. For a detailed presentation of the De Broglie-Bohm interpretation, refer to \cite[]{bohm1981wholeness} and \cite[]{holland1995quantum}. 

\subsection*{Level 1 - The mathematical formalism}

\begin{eqnarray}
\psi: \mathbb{R}^{3N} \times \mathbb{R} \to \mathbb{C} \\
\psi=\psi(\vec q_k, t) \\
k = 1,..., N \\
\psi=R e^{i\frac{S}{\hbar}} \\
\label{equ_Polar decomposition}
R: \mathbb{R}^{3N} \times \mathbb{R} \to \mathbb{R} \\
S: \mathbb{R}^{3N} \times \mathbb{R} \to \mathbb{R} 
\end{eqnarray}

\begin{equation}
i\hbar \frac{\partial \psi}{\partial t}(\vec q,t)=-\sum_{k=1}^{N} \frac{\hbar^2}{2m_k}\Delta_k \psi(\vec q,t)+V(\vec q)\psi(\vec q,t) 
\label{eqn_Schrödinger1}
\end{equation}

\begin{equation}
\frac{\partial S}{\partial t} + \sum_{k=1}^{N}  \frac{(\nabla_k S)^2}{2m_k}+\sum_{k=1}^{N} - \frac{\hbar^2\nabla^2_k R}{2m_kR}+V=0
\label{equ_Real Hamilton Jacobi}
\end{equation}

\begin{equation}
\frac{\partial (R^2)}{\partial t} + \sum_{k=1}^{N} \nabla_k \cdot \frac{R^2\nabla_k S}{m_k} =0
\label{equ_Imag_Conituity equation}
\end{equation}

\subsection*{Level 2 - Physical quantities and connection to empirical rules}

The mathematical level contains the Schr\"odinger equation. Every complex valued function can be written in a polar form of the kind in equation (\ref{equ_Polar decomposition}) with two real valued functions, $R$ and $S$. Applying the polar form on the Schr\"odinger equation, one obtains equations (\ref{equ_Real Hamilton Jacobi}) and (\ref{equ_Imag_Conituity equation}). This is just a reformulation of the Schr\"odinger equation. \newline

Interpreting equation (\ref{equ_Real Hamilton Jacobi}), one sees that it resembles the Hamilton-Jacobi equation of classical mechanics. A detailed account of Hamilton-Jacobi theory in the context of the De Broglie-Bohm interpretation is provided by \cite[p.27]{holland1995quantum}. The term $S$ resembles the action of classical mechanics, but there is an additional term that has no counterpart in classical Hamilton-Jacobi mechanics; it is the term we will denote by $Q$ and is usually named Quantum potential:

\begin{equation}
Q = \sum_{k=1}^{N} - \frac{\hbar^2\nabla^2_k R}{2m_kR}
\label{equ_Quantum potential}
\end{equation}

The Quantum potential is interpreted on the same footing as a potential energy, but due to its structure, it has some specific properties that are emphasized as in-formational \cite[p.89]{holland1995quantum}:

\begin{quote}
\textit{``Thus contary to what one might expect in a classical wave, a particle does not respond to the intensity of the wave in its vicinity, but rather to its form.''}
\end{quote}

The proponents of the De Broglie-Bohm interpretation often make use of a radio-wave analogy for the Quantum potential and the in-formational character. The radio waves that control the movement of a remote-controlled vehicle carry an in-formational content that is independent of the amplitude of the radio wave. Essential is its informational content that guides the remote controlled object. In the sense needed on Level 2 we interpret the term $Q$ as a potential, equally important to the guidance of the particle as the classical potential §$V$. \newline

By interpreting equation (\ref{equ_Real Hamilton Jacobi}) as a Hamilton-Jacobi type equation in the De Broglie Bohm interpretation, one concludes that there are equations of motions for the individual systems \cite[]{holland1995quantum}:

\begin{quote}
\textit{``... let us construct a vector field ... and assume that the latter defines at each point of space at each instant the tangent to a possible particle trajectory passing through that point. This naturally provides a description of an ensemble of particles, which is fictitious in the sense that only one track is realized in any given field ...''}
\end{quote}

\newpage

For one particle, this vector field is given by

\begin{equation}
\frac{d\vec q}{dt} = \frac{\vec p}{m} = \frac{\nabla S}{m}.
\end{equation}

This equation is called guidance equation and provides a set of trajectories as solutions. In a many-particle situation, the guidance equation is given by 

\begin{equation}
\frac{d\vec q_k}{dt} = \frac{\nabla_k S}{m_k}.
\end{equation}

The $\vec q_k(t)$ represent the positions of the particles at specific times $t$. \newline

The interpretation of the function $R$ is consistent with the Born interpretation in some aspects, but not in all \cite[p.325]{bohm1987ontological}: In that sense equation (\ref{equ_Imag_Conituity equation}) ...

\begin{quote}
\textit{``... can evidently be regarded as a continuity equation with $P = R^2$ being a probability density, as Born suggested. The function $P$ has, however, two interpretations, one through the Quantum Potential and the other through the probability density. It is our proposal that the fundamental meaning of $R$ (and therefore indirectly of $P$) is that it determines the Quantum Potential. A secondary meaning is that it gives the probability density for the particle to be at a certain position. Here we differ from Born who supposed that it was the probability of finding the particle there in a suitable measurement.''}
\end{quote}

\subsection*{Level 3 - Concepts and principles}

\paragraph{Particle and wave conception:}

A particle lives in the $\mathbb{R}^3$ and is defined by its particle trajectory $q_k(t)$. A particle is guided by the wave function or  two real valued components $R$ and $S$ \cite[p.277]{holland1995quantum}:

\begin{quote}
\textit{``When we speak of a "many body system" we mean then a single wavefunction together with a set of particles. There is in general no wave associated just with each particle individually.''}
\end{quote}

It is important to emphasize this aspect of the many particle situation in the De Broglie-Bohm interpretation, as an essential point gets lost if one studies only the single particle situation. In the single particle case, the configuration space is a $\mathbb{R}^3$ and isomorphic to the physical space $\mathbb{R}^3$. This gives the impression that the De Broglie-Bohm interpretation is acting on the physical space. In fact, it does, but only the positions $q_k(t)$ refer to physical space, and the functions $R$ and $S$ and therefore the $\psi$ live on the $\mathbb{R}^{3N}$ configuration space. 

In the De Broglie-Bohm interpretation, the wave function has no primary conceptual status; it is decomposed in the two real valued functions $R$ and $S$. Nevertheless, the De Broglie-Bohm interpretation refers also to the wave function and outlines characteristics. We follow the presentation by \cite[p.83-84]{holland1995quantum} and present those aspects relevant to our discussion:

\begin{itemize}
\item
There is no source of the $\psi$-field. It has no localized origin.

\item
There is no ontological construction for a medium where the $\psi$-function oscillates on. 

\item
Due to the formal construction of the Quantum potential influences by the $\psi$-function are independent of the intensity and therefore nonlocalities are easily realised.

\item
Due to the formal construction of the Quantum potential, influences by the $\psi$-function are independent of the intensity of the $\psi$-function respectively of the function $R$ and therefore nonlocalities are easily realised.

\end{itemize}

\paragraph{Hidden variables:}

The De Broglie-Bohm interpretation is a hidden variable theory; these hidden variables are the particles positions $\vec q_k(t)$and their initial values $\vec q_o$.  

\paragraph{Quantum potential:}

One conception is emphasized in the De Broglie-Bohm interpretation. It is the conception of the Quantum potential represented by equation (\ref{equ_Quantum potential}). The Quantum potential plays the key role role in the conceptual apparatus of the the De Broglie-Bohm interpretation \cite[p.78]{holland1995quantum}:

\begin{quote}
\textit{``The introduction of the quantum potential as a causal agent has explanatory power which one unnecessarily forgoes by concentrating on just (3.2.19)\footnote{The guidance equation.}. It represents the difference between classical and quantum mechanics.''}
\end{quote}

Though proponents of the De Broglie-Bohm interpretation claim a resemblance of the mathematical apparatus with classical mechanics, particularly Hamilton-Jacobi mechanics, they do not aim for a return to classical conceptions or ontology. The Hamilton-Jacobi formalism is merely a unified language that speaks about phenomena in classical and quantum context \cite[p.78]{holland1995quantum}:

\begin{quote}
\textit{``The radical departure from classical notions inherent in the causal interpretation\footnote{The De Broglie-Bohm interpretation is often referred to as the causal interpretation.} lies in the details of the model, in the types of motions that are accessible to particles.''}
\end{quote}

\newpage

\subsection*{Level 4 - Ontology}

\paragraph{Particle and wave ontology:}

The particle ontology of the De Broglie-Bohm interpretation leads to a new conception of matter \cite[p.65]{holland1995quantum}:

\begin{quote}
\textit{``The classical theory of matter, ... has been replaced by a more general conception in which matter has an intrinsic field aspect, the mass points moving and interacting under the influence of the new internal energy as well as the more familiar potentials of classical dynamics.
This leads to a synthesis of the wave and particle characteristics of matter. These properties are not mutually exclusive, but simultaneously existent. ''}
\end{quote}

A particle is not only an entitiy represented by a set of properties and a position. In the De Broglie-Bohm interpretation, the particle conception goes beyond that classical naive representation of particles \cite[]{bohm1987ontological}:

\begin{quote}
\textit{``As the theory develops, we shall find that the electron is by no means a structureless particle. Rather, what is suggested by its behaviour is that it is a highly complex entity that is deeply affected by its quantum field in an extremely subtle and dynamic way. Moreover, this entity is not to be regarded (as is done in the usual interpretations) as somehow directly possessing both particle-like and wave-like properties. Rather, the observed wave-like properties will follow, as we shall see, from the general effect of the quantum wave field on the complex structure of the particle. ''}
\end{quote}

The particle wave ontology is not a simple replacement of the classical wave and particle pictures by an assertion that quantum systems behave in some situations wave-like and in some situations particle-like. It is more; it is the view that matter behaves in a novel way in the quantum regime that is incompatible with both ontological pictures of particle and waves alike. On the question, what are the quantum objects, an appropriate answer would be: something different.

\paragraph{Active information and the causa formalis:}

The De Broglie-Bohm interpretation emphasizes a conception of in-formation that is related to the Aristotelian causa formalis \cite[p.16]{bohm1981wholeness}: 

\begin{quote}
\textit{``In more modern language, it would be better to describe this as formative cause, to emphasize that what is involved is not a mere form imposed from without, but rather an ordered and structured inner movement that is essential to what things are''}
\end{quote}

By proponents of the De Broglie-Bohm interpretation this term was used to describe this feature in sense of an in-formation: Active information. \cite[p.19]{hiley2005can} describes its features: 

\begin{quote}
\textit{``The proposal is that active information at the quantum level organizes
the dynamical evolution of the system itself.
...
As we remarked earlier, the quantum potential appears to be some kind of internal energy which carries information about the environment. Therefore we consider the idea that the whole process, particle plus active environment (which in terms of a measurement requires the specification of experimental conditions), is being
formed partly from within. This suggests that the process is more organic than mechanical.''}
\end{quote}

Active information can be understood in the sense of the radio-wave remote-controlled vehicle. The vehicle's motions depend not only on its own local laws, but on some information on the environment provided by a formative cause. 

A recent account on the role of the Quantum potential in the context with Active information is provided by \cite[]{dennis2015bohm}.

%
%
%

\paragraph{Ontological Holism:}

The De Broglie-Bohm interpretation advances an ontological holism. The Quantum potential reflects this holism and is the element in the theory that acts in the sense of this wholeness \cite[p.340]{bohm1987ontological}: 

\begin{quote}
\textit{``We have seen thus far that the quantum behaviour of matter shows a certain kind of wholeness, brought about by the quantum potential. This latter functions as active information that may reflect distant features of the environment and may give rise to a non-local connection between particles that depends on the “quantum state” of the whole, in a way that is not expressible in terms of the relationships of the particles alone.''}
\end{quote}

According to the ontology advanced by the De Broglie-Bohm interpretation, there is a wholeness in the very nature of reality that is hidden to us in some branches of phenomena, like classical physics, but revealed by Quantum theory. Wholeness in that respect affects also the weight and relevance of our scientific theories \cite[p.21-p.22]{bohm1981wholeness}:

\begin{quote}
\textit{``We thus to be alert to give careful attention and serious consideration to the fact that our theories are not descriptions of reality as it is, but rather ever-changing forms of insight, which can point to or indicate a reality that is implicit and not describable or specifiable in its totality.''}
\end{quote}

\newpage

\subsubsection*{Overview} 
 
 We present the De Broglie-Bohm interpretation structure according to the Classification scheme in an overview. Differences from Bohmian Mechanics, which will be presented in section \ref{section Bohmian Mechanics}, are discussed in the conclusion.
 
 \begin{table}[h]
 \centering
 
 \label{SchemedBB}
 \begin{tabular}{|l|c|}
 \hline
 \begin{tabular}[c]{@{}c@{}}\\ 1 \\ \\ \end{tabular}  & \begin{tabular}[c]{@{}c@{}} \\  $\frac{\partial S}{\partial t} + \sum_{k=1}^{N}  \frac{(\nabla_k S)^2}{2m_k}+\sum_{k=1}^{N} - \frac{\hbar^2\nabla^2_k R}{2m_kR}+V=0$ \\ \\ $\frac{\partial (R^2)}{\partial t} + \sum_{k=1}^{N} \nabla_k \cdot \frac{R^2\nabla_k S}{m_k} =0 $ \\ \\ \end{tabular} \\   \hline
 \begin{tabular}[c]{@{}c@{}}\\ 2 \\ \\ \end{tabular} & \begin{tabular}[c]{@{}c@{}}  \\ $\vec q_k(t)$... position of an object at time $t$ \\ $m_k$ ... mass of a  particle \\ $S$ ... Hamilton Jacobi function or action \\ $R^2$ ... probability \\ $Q$ ... Quantum potential \\ \\ Measurement rules for $\vec q_k(t), m_k, R^2 $  \\  \\ \end{tabular} \\ \hline
 \begin{tabular}[c]{@{}c@{}}\\ 3 \\ \\ \end{tabular} & \begin{tabular}[c]{@{}c@{}} \\ Particle and wave conception \\ Hidden variables \\ Quantum potential \\ \\ \end{tabular}                                                                                 \\ \hline
 4 & \begin{tabular}[c]{@{}c@{}}\\ Novel matter ontology \\ Active information \\ Wholeness and ontological holism \\ \\ \end{tabular}                      \\ \hline
 \end{tabular}
 \caption{Classification scheme of the De Broglie-Bohm interpretation}
 \end{table}

 \newpage
\subsection{Bohmian Mechanics}
\label{section Bohmian Mechanics}
Bohmian Mechanics is a reconstruction of Quantum Mechanics based on the ideas of the De Broglie-Bohm interpretation, but postulates the guidance equation for particle positions and not as a consequence of the Schr\"odinger equation. It reproduces the results and experimental predictions of standard Quantum Mechanics, and is a realistic, deterministic, nonlocal hidden-variable theory. Its fundamental ontological entities are the particles and their position. We analyse Bohmian Mechanics for N-particle systems without spin, because one-particle systems easily give a wrong impression on the nature of Bohmian Mechanics. Main proponents of Bohmian Mechanics are Detlef D\"urr, Sheldon Goldstein and Nino Zanghi. For further reading, see \cite[]{durr2009bohmian} and \cite[]{cushing2013bohmianmech}. 

\subsection*{Level 1 - The mathematical formalism}

The mathematical level consists of two equations: the Schr\"odinger equation and the guidance equation. The function $\Psi$ is a mapping from a $\mathbb{R}^{3N} $ vector space to the complex numbers together with the free parameter $t$.  \newline

\begin{eqnarray}
\psi: \mathbb{R}^{3N} \times \mathbb{R} \to \mathbb{C} \\
(\vec q,t) \mapsto \psi(\vec q,t) 
\end{eqnarray}

\begin{eqnarray}
\vec q = \left( \begin{array}{c} \vec q_1 \\ . \\ . \\ . \\ \vec q_N \end{array} \right) \in \mathbb{R}^{3N}
\end{eqnarray}

\begin{equation}
i\hbar \frac{\partial \psi}{\partial t}(\vec q,t)=-\sum_{k=1}^{N} \frac{\hbar^2}{2m_k}\Delta_k \psi(\vec q,t)+V(\vec q)\psi(\vec q,t) 
\label{eqn_Schrödinger2}
\end{equation}

\begin{equation}
\frac{d\vec Q_k}{dt}=\frac{\hbar}{m_k}\Im \frac{\nabla_k\psi}{\psi}(\vec Q,t) 
\label{eqn_Leitgleichung}
\end{equation}

\newpage

\subsection*{Level 2 - Physical quantities and connection to empirical rules}

Many-particle Bohmian Mechanics assigns properties to the $k-th$ particle\footnote{A conception that will be defined on level three}, its mass $m_k$, and its trajectory represented by a vector $\vec Q_k(t) \in \mathbb{R}^3 $. This trajectory refers to the empirical accessible positions of particles\footnote{In the sense of the empirical phenomena.} in time. \newline

The laws of motion for $N$ particles are formulated in a $\mathbb{R}^{3N}$ configuration space, where the complex valued $\psi$ function is defined. The $\psi$ function obeys equation (\ref{eqn_Schrödinger2}). A solution can be provided when the function $V(\vec q)$ is defined. $V(\vec q)$ represents the physical potentials the particles are exposed to. The $\psi$ function is called wave function.

The trajectories of the particles obey equation (\ref{eqn_Leitgleichung}), which defines the trajectories due to the wave function, which is itself a solution of equation (\ref{eqn_Schrödinger2}). The initial conditions of $Q_k(t)$ and $\frac{dQ_k(t)}{dt}$ have to be defined to find unique solutions for the trajectories. \newline

The statistical distributions of quantum systems are reproduced by Bohmian Mechanics; specifically, by the Quantum Equilibrium hypothesis \cite[p.153]{durr2009bohmian}: 

\begin{quote}
\textit{``Quantum Equilibrium Hypothesis. For an ensemble of identical systems, each having the wave function $\psi$, the typical empirical distribution of the configurations of the particles is given approximately by $\rho=|\psi|^2$. In short, Born´s statistical law holds.''}
\end{quote}

This is a brief summary of the most important definitions and conceptions within Bohmian Mechanics. Some aspects of the physical formalism presented here follow out of a principle, the already introduced Quantum Equilibrium hypothesis.  

\newpage

\subsection*{Level 3 - Concepts and principles}

Quantum equilibrium hypothesis: The quantum equilibrium hypothesis has the status of a concept or first principle in Bohmian Mechanics. \newline

The empirical statistical distributions of quantum systems are reproduced by Bohmian Mechanics; specifically by the Quantum Equilibrium hypothesis. It can be shown by Bohmian Mechanics that if a quantum system's positions are distributed according to Born's statistical interpretation, then they will stay in this quantum equilibrium. The evolution of the $\psi$ function maintains the quantum equilibrium \cite[p.211ff]{durr2009bohmian}.

\paragraph{Particle and wave conception:}

Particle: A particle is represented by a trajectory $\vec Q_k(t)$ and lives in the $ \mathbb{R}^3$. \newline

The wave assigned to Quantum Theory is represented by the wave function $\psi$ and lives in the configuration space $ \mathbb{R}^{3N}$. 

\paragraph{Hidden variables:}

Bohmian Mechanics is a hidden variable theory; these hidden variables are the particle's positions and their initial values.  

\subsection*{Level 4 - Ontology}

\paragraph{Particle Ontology:}

Bohmian Mechanics advances a strict particle ontology. The ontological objects are the particles, some of their properties and their positions \cite[p.142]{durr2009bohmian}: 

\begin{quote}
\textit{``Bohmian Mechanics will ecplain that it is correct to say "is" for the positions of the systems particles, but that it is not correct to say "is" for other "observables".''}
\end{quote}

D\"urr alludes to the fact that not all observables have an ontological status. In fact, position plays a distinguished role. Momentum or spin are not on the same ontological footing. They are not properties of the particles, the ontological objects themselves. 

A further aspect of the particle ontology is that it is an object ontology. There are objects out there; these objects are the particles in a physical real space. Referring to the structuralist approaches where structures play an important role, moderate ontological structural realism or even more epistemological structural realism would be philosophical stances that are compatible with Bohmian Mechanics. 

All particles are guided by the wave function simultaneously, but this wave function lives in the configuration space, which has a dimension depending on the number of particles involved, or if we assume the wave function of the universe, it would have a dimension according to the number of particles in the universe. But it is not that easy to abandon the ontological footing of the wave function \cite[p.5-6]{esfeld2013ontology}:

\begin{quote}
\textit{``There is indeed a prima facie good reason to do so and to admit the wave-function as a further concrete physical entity in addition to the particles. ''}
\end{quote}

Mainly, the reason for this is grounded on the causal relation of the wave function to the movement of the particles. The aim for such a causal connection that the wave function is a guiding field led De Broglie to propose the original version of the guidance conception for the wave function. A further problem of this ontology would be \cite[p.7]{esfeld2013ontology}:

\begin{quote}
\textit{``However, if one admits configuration space as a further stage of physical reality – in addition to and independent of three-dimensional physical space –, it is unclear how there could be a real connection between these two spaces that could amount to something existing in the one space guiding or piloting the motion of entities existing in the other one. ''}
\end{quote}

The possibilities of solving this ontological connection range from neglecting the idea of any ontological footing of the wave function, overassigning the wave function to  physical space, to the position that configuration space itself represents a fundamental ontology \cite[p.7]{esfeld2013ontology}.

Two positions seem promising and represent an ontological dualism within Bohmian Mechanics: Humeanism and Dispositionalism \cite[p.25]{esfeld2013ontology}: 

\begin{quote}
\textit{``...Humeanism – more precisely, Lewis’ thesis of Humean supervenience: the world according to quantum mechanics can be “a vast mosaic of local matters of particular fact”, namely the spatio-temporal distribution of the elements posed in the primitive ontology such as particle positions. The universal wave-function and the laws of quantum mechanics supervene on this distribution. They are nothing more than devices of economical bookkeeping, there being no real connections among the elements of the primitive ontology... ''}
\end{quote}

\cite[]{esfeld2013ontology} emphasize that the mere fact of the the existence of an approach like Bohmian Mechanics refutes the view that Quantum Mechanics is in contradiction with Humeanism. In this view, the wave function is a instrumentalistic element of the theory without any footing in ontology. It is helpful to describe the situation but gives no deeper insight in the nature of reality. A classical analogy would be the center of mass of a bulk of matter. The center of mass conception is very helpful, but no ontological speciality is located at the position of the center of mass of the bulk of matter. It is just a nomological entity. \newline

Dispositionalism, on the other hand, is the view that there is an inherent property of local beables, the particles, that ensures their evolution in a specific way. To \cite[]{esfeld2013ontology} it brings the advantage that: 

\begin{quote}
\textit{``... one can include something in the ontology that accounts for the temporal
development of the local beables, without having to resort to accepting the
wave-function in configuration space as an element of physical reality.''}
\end{quote}

\cite[]{esfeld2013ontology} favour dispositionalism but admit that it is an open research project to clarify the eligibility in foundation of physics:
\begin{quote}
\textit{``On the question of Humeanism versus dispositionalism we side with dispositionalism, since we take it to be a sound demand to call for something in the ontology that accounts for the temporal development of the elements of physical reality and that grounds the law of motion, thus providing for real connections in nature.''}
\end{quote}

\paragraph{Nonlocality and Holism:}

Bohmian Mechanics is a nonlocal theory and its ontology is also nonlocal and in some sense holistic \cite[p.131]{durr2009bohmian}: 

\begin{quote}
\textit{``What is new\footnote{In Quantum Mechanics compared to classical physics.} is that the description of nature needs a function on the configuration space of all particles in the system. And why is that revolutionary? The point is that such a description involves all particles in the universe at once, whence all particles are "entangled" with each other,...''}
\end{quote}

The Nonlocality in Bohmian Mechanics reflects an ontological holism as proposed in the De Broglie-Bohm interpretation, but Bohmian Mechanics puts no emphasis on this aspect and takes it merely as a consequence of the mentioned dispositions of the evolution of the local beables. 

\newpage

\subsection*{Overview}

We apply the classification scheme to give an overview on the structure of Bohmian Mechanics. Significant differences concerning all levels of the classification scheme between the De Broglie-Bohm interpretation and Bohmian Mechanics are evident.

\begin{table}[h]
\centering

\label{SchemeBohmianMechanics}
\begin{tabular}{|l|c|}
\hline
\begin{tabular}[c]{@{}c@{}}\\ 1 \\ \\ \end{tabular}  &  \begin{tabular}[c]{@{}c@{}}  \\ $i\hbar \frac{\partial \psi}{\partial t}(\vec q,t)=-\sum_{k=1}^{N} \frac{\hbar^2}{2m_k}\Delta_k \psi(\vec q,t)+V(\vec q)\psi(\vec q,t) $  \\  \\ $\frac{d\vec Q_k}{dt}=\frac{\hbar}{m_k}\Im \frac{\nabla_k\psi}{\psi}(\vec Q,t) $ \\   \\ \end{tabular}  \\  \hline
\begin{tabular}[c]{@{}c@{}}\\ 2 \\ \\ \end{tabular} & \begin{tabular}[c]{@{}c@{}}  \\ $\vec Q_k(t)$... trajectory of the $k-th$ particle  \\ $m_k$ ... of the $k-th$ particle \\ $\psi(\vec q,t) $ ... wave function in the $\mathbb{R}^{3N} $ \\ $\rho=|\psi|^2$... Born's statistical interpretation \\ \\ Measurement rules for $Q_k(t), m_k, \rho $  \\  \\ \end{tabular} \\ \hline
\begin{tabular}[c]{@{}c@{}}\\ 3 \\ \\ \end{tabular} & \begin{tabular}[c]{@{}c@{}} \\ Point particle concept \\ Quantum equilibrium hypothesis \\ \\ \end{tabular}                                                                                 \\ \hline
4 & \begin{tabular}[c]{@{}c@{}}\\ Particle ontology \\ Position in  $\mathbb{R}^3 $ space and time \\ Wave ontology: Humeanism vs Dispositionalism \\ \\ \end{tabular}                      \\ \hline
\end{tabular}
\caption{Classification scheme of Bohmian Mechanics}
\end{table}

\newpage

\section{Conclusion}

We presented a classification scheme to provide a tool to distinguish interpretations of Quantum Mechanics concerning their conceptions, formalism and ontology. The classification scheme consists of four levels, which refer to the mathematical formalism, the interpretation of the formalism, the used concepts and the ontological level. The first three refer to the theory; the latter contains the metaphysical implications of the theory. \newline

In philosophy of science, a manifold field of conceptual approaches on the structure of physical theories developed in the last century. We presented the relation of these approaches to our introduced classification scheme. We started with Einstein's model of layers of a scientific theory that represented only the progress of a scientific theory in respect to its formal most simple representation. With the advent of Logical Empiricism positions on the structure of scientific theories also developed. We presented the received view and its successors the semantic view and the structuralist approaches. All of these positions have to be seen in the dichotomy of realist and anti-realist positions.    \newline

Out of the structuralist approaches, the position of structural realism evolved, which developed directly related to Quantum Mechanics and the debates on the foundations of its interpretation. \newline

Concerning the ontological level of a physical theory, we proposed two arguments that refer to the interrelation between two or more physical theories and to the relation between theoretical terms and ontological truth. \newline

The Ontological Coherence Argument is a heuristic that provides advice as to which theories should be adopted based on their ontological relations to other theories.  \newline

Concerning the truth values of scientific theories, we provided a theorem on ontological truth, which states that scientific knowledge is something that is cumulative on one hand, but not convergent on the other hand. The theorem reflects the progress of science by acknowledging the value of exclusive statements and their contribution to the growth of scientific knowledge. \newline

We propose that the Ontological Coherence Argument is applied implicitly by proponents of a realist point of view in the interpretations of Quantum Mechanics when they adopt a Bohmian-type or a Ghirardi-Rimini-Weber-position. A realist view in the sense of a pre-existing framework of space and time and properties of objects, as realised by theory of relativity, is supported by both positions in Quantum Mechanics. To proponents of a realist point of view, as in the mentioned interpretations, this coherence of different theories is a strong supporting argument. 

Since the interpretation of space-time is an open question in philosophy of science as well, the Ontological Coherence Argument can provide support here as a useful heuristic by relating interpretational problems of Quantum Mechanics with interpretational problems of space-time. Structural realism is applied in both fields, and is a promising conception for solving interpretational questions. \newline

The relation of these two arguments to scientific underdetermination and the unreasonable effectiveness of mathematics remains an open question and is subject to future work. \newline


We demonstrated the classification scheme by means of classical particle mechanics, and applied it to the field of interpretations of Quantum Mechanics by the example of the De Broglie-Bohm Interpretation and Bohmian Mechanics. The classification scheme demonstrates the differences in the conception, the mathematical formalism and the ontology of these two interpretations of Quantum Mechanics. In the case of Bohmian Mechanics and the De Broglie-Bohm Interpretation, the classification scheme demonstrated the profound differences in the concepts and ontology of these two approaches. Whereas Bohmian Mechanics emphasizes a particle conception and ontology, the De Broglie-Bohm Interpretation emphasizes holism and an abandoning of reductionism. We demonstrated that despite, in literature, the two approaches are mixed up or seen as only one approach, the conceptions and ontologies are radically different. \newline

The application of the classification scheme to other interpretations of Quantum Mechanics can be done in a straightforward way and is subject to future work.  

\section{Acknowledgments}

I would like to thank the head of the Quantum Particle Workgroup, Beatrix Hiesmayr, for her support and Basil Hiley, Chris Fuchs, Stefanie Lietze, and several other colleagues for fruitful discussions and comments. For proofreading I would like to thank Daniel Moran. For the realisation of the graphical elements and illustrations, I would like to thank Isabella W\"ober and Daniel Alpar.

\end{document}